\documentclass{article}
\usepackage{graphicx}

\author{
   Niels G. Gresnigt\footnote{Corresponding author}\\
   \texttt{niels.gresnigt@xjtlu.edu.cn}\\
  \small{Department of Mathematical Sciences}\\
  \small{Xi'an Jiaotong-Liverpool University}\\
  \small{111 Ren'ai Road, Suzhou HET, Jiangsu}\\
  \small{China 215123}\\
  \and 
  Adam B. Gillard\\
  \texttt{adam.gillard@pau.ac.pg}\\
  \small{School of Science and Technology}\\
  \small{Pacific Adventist University}\\
  \small{14 Mile Sogeri Road, NCD,} \\
  \small{Papua New Guinea 111}
}

\title{Electroweak symmetries from the topology of deformed spacetime with minimal length scale}
\begin{document}
\maketitle

\begin{abstract}
Lie-type deformations provide a systematic way of generalising the symmetries of modern physics. Deforming the isometry group of Minkowski spacetime through the introduction of a minimal length scale $\ell$ leads to anti de Sitter spacetime with isometry group $SO(2,3)$. Quantum spacetime on scales of the order $\ell$ therefore carries negative curvature.

Considering extended particles of characteristic size $\ell$ carrying topological information and requiring that their topological properties be compatible with those of the underlying spacetime, we show that electroweak symmetries emerge from the maximal compact subgroup of the anti de Sitter isometry group in a way that is consistent with no-go theorems. 

It is speculated that additional deformation outside the Lie-algebraic framework, such as $q$-deformations, could likewise provide an explanation of the origin of the strong force.

\end{abstract}
\section{Introduction}

Lie-algebraic stability provides an important physical principle. The numerical values of physical constants in theories are not determined algebraically but must be measured experimentally. Consequently these constants are not known without some degree of uncertainty. One should therefore search for theories that do not depend critically on the exact numerical values of its parameters. Physical theories which do not change in a qualitative matter under a small change of the parameters are said to be physically {\it stable}. An unstable theory should be deformed until a stable theory is reached which is likely to be of wider validity than that of the original unstable theory.
 


Much of modern physics including quantum field theory and the Standard Model (SM) has been formulated through the spacetime symmetries of special relativity, encoded within the Poincar\'e algebra, together with the principle of local gauge invariance. The Poincar\'e algebra is sensitive to small perturbations in its structure constants however and therefore not stable.



Despite its success in accurately describing particle interactions via the electroweak and strong nuclear force, the SM remains an incomplete theory with a number of peculiarities that hint at the existence of a more fundamental theory from which the SM arises. Many attempts to explain the peculiar features of the SM, such as the large number of arbitrary parameters, are based on replacing point particles with extended topological objects such as knots of electromagnetic flux, or braids. 

At the same time, given the instability of the Poincar\'e algebra, its validity as underlying the SM symmetries is questionable. We argue that a natural path toward a more widely applicable and robust theory of elementary particles and their interactions is to replace the underlying unstable algebra by a larger and stable deformed one.  

There are two paths (within our Lie-algebraic setting) along which to proceed in our attempt to generalize the isometries of quantum relativity. The first of these is via stabilization of the combined Poincar\'e-Heisenberg algebra, consisting of the Poincar\'e algebra together with the position operators and Heisenberg commutation relations. This has been the approach of Mendes \cite{mendes1994,mendes2012deformation}, Chryssomalakos and Okon \cite{chryssomalakos2004:1}, and Ahluwalia \textit{et. al.} \cite{ahluwaliakhalilova2005:1,ahluwalia2008ppa}. 

The approach taken here however is to restrict ourselves to the momentum sector, and look for deformations of the Poincar\'e algebra only. This has also been the approach adopted by Kong and his research group \cite{das2006physics,kong2007quantum,kong2008deformed,kong2009ads_5,kong2009poincare,cho2014relativity}. In the general case, both approaches give rise to two additional deformations (with two associated invariant scales) with the resulting semisimple, and hence stable group being $SO(2,4)$  \footnote{In somewhat more generality, the approach of stabilizing the combined Poincar\'e -Heisenberg algebra leads to the group $SO(m,6-m)$. The specific value of $m$ is then dependent on the identification of the deformation parameters with specific invariant scales.}, which is stable and can not be deformed further. 

The internal symmetries described in the SM, contained in the gauge group $SU(3)\times SU(2)\times U(1)$, presents one of the most controversial aspects of the model as the fundamental origin of the internal symmetries is not known. By replacing point particles with extended objects carrying topological information as well as requiring the isometries of spacetime to be stable, we find that the electroweak symmetries emerge from the topological information carried by the underlying deformed spacetime in a way that is consistent with no-go theorems. Our derivation does not require us to assume the specific topological nature of elementary particles but rather only that it be consistent with the topology of the underlying spacetime in which they live.


The approach taken in this paper is guided by two underlying principles. The first is that one should look for stable Lie algebras and isometry groups to describe the spacetime symmetries underlying quantum field theory and high energy physics. Our second guiding principle is that the topological properties of elementary particles must be compatible with the topological information of the underlying spacetime. This second principle is based on the hypothesis that internal symmetries might be topological in origin. 

In Galilean relativity all kinematic variable such as speed and momentum are unconstrained and can take any real value. Restricting velocities by imposing a maximum speed $c$ deforms the space(time) metric and its isometries. The resulting isometries are those encoded within the Poincar\'e algebra. Imposing additional kinematic constraints deforms the metric and isometries further. The minimal requirement for deforming the Poincar\'e isometries into a stable isometry group is through the introduction of an invariant minimal quantum length scale $\ell$ in addition to the speed of light\footnote{We note that it is possible to introduce additional invariant scales through repeated stabilizations and addition of translations groups to the isometry group which renders the full isometry group unstable again. This process terminates once a length scale is introduced. In the present paper we consider the minimal possible deformation which corresponds to introducing the length scale straight way. This is also the minimal deformation required to constrain all of the kinematic variables.}. Together, $\ell$ and $c$ provide a full set of constraints for all kinematic variables including a maximum speed, momentum and energy, and a minimum length and time.

The emerging picture of spacetime at the quantum scale as a result of deforming the Poincar\'e algebra, via a minimal quantum length scale, is that of anti de Sitter space AdS with isometry group $SO(2,3)$\footnote{Experimental observations suggest the universe on cosmological scales carries a positive curvature meaning a positive cosmological constant and the isometries of de Sitter rather than anti de Sitter space. Our arguments do not contradict this view of spacetime. Rather we are saying that on the quantum scale, spacetime looks like anti de Sitter space. In a future paper we address this issue in greater detail.}. 

We comment briefly on the irreducible representations of the quantum spacetime isometry and the new physics it leads to. In particular we speculate that the singleton representations of $SO(2,3)$ may be related to various preon models of particles but leave an in-depth investigation for a future time.

The homotogy groups, including the fundamental group, of anti de Sitter space are the same as those of its maximal compact subgroup. Requiring compatibility between the topological properties of particles and that of the underlying spacetime suggests the maximal compact subgroup to play and important role in the description of particles. We demonstrate that the electroweak gauge group is locally isomorphic to the maximal compact subgroup of the anti de Sitter isometry group. We speculate that further deformations, outside the Lie-algebraic framework, may lead to an understanding of the strong force.
\section{The principle of Lie-algebraic stability}\label{PLAS}

A Lie algebra is said to be stable if small perturbations in its structure constants result in isomorphic algebras. Stability of an algebra is equivalent to saying a physical theory based on the algebra is robust and free of fine tuning issues. This concept of Lie-algebraic stability provides insight into the validity of a physical theory or the need to generalize the theory. An unstable theory might be deformed until a stable theory is reached which is likely to be of wider validity than that of the original unstable theory. For details about the theory and mathematical process of stabilizing a Lie algebra, the reader is directed to Gerstenhaber \cite{gerstenhaber1964}, Nijenhuis and Richardson \cite{nijenhuis1967}, and Chryssomalakos and Okon \cite{chryssomalakos2004:1}. 

In the present paper we restrict ourselves to deformations of Lie-type for two reasons. First, Lie algebras have played a fundamental role in many successful physical theories; and second, the mathematical formalism of Lie-algebraic deformations is well developed and thus deformations of this kind can be handled systematically \footnote{In future work we intend to relax this restriction to Lie-type deformations and also consider $q$-deformations \cite{lukierski1991q,lukierski2003doubly,lukierski2002doubly,majid1994bicrossproduct}. These types of deformations, like Lie-type deformations, have been well studied and can likewise be handled systematically.}. 

The importance of Lie-algebraic stability in physical theories was first promoted by Mendes \cite{mendes1994}. Since then, several others, most notably Chryssomalakos and Okon \cite{chryssomalakos2004:1}, and Ahluwalia \cite{ahluwaliakhalilova2005:1} have similarly argued that the stability of a physically relevant Lie algebra should be considered a physical principle, on an equal footing as the principle of local gauge invariance, and provides a necessary algebraic requirement (although perhaps not a sufficient requirement) for a Lie algebra to represent physics that is invariant under imperfect measurements.

Faddeev \cite{faddeev1989} was the first to point out (albeit in hindsight) that both the quantum and relativistic revolutions of the 20th century can be considered as Lie-algebraic stabilizations of the algebras of classical mechanics and Galilean relativity respectively. The stabilization of these algebras give the Heisenberg and Lorentz algebras, which are both individually stable. Mendes further noticed that although the Lorentz and Heisenberg algebras are separately stable, the combined Poincar\'e-Heisenberg algebra lacks the desired stability \cite{mendes1994}.

Following his observation that the combined Poincar\'e-Heisenberg algebra is not stable, Mendes proceeded to find its stabilized form. The algebra turned out to be (up to various signs) the same algebra arrived at by Yang \cite{yang1947} in 1947 based on the work of Snyder who demonstrated that the assumption that space be a continuum is not required for Lorentz invariance earlier the same year \cite{snyder1947}. Uniqueness was later demonstrated by Chryssomalakos and Okon \cite{chryssomalakos2004:1}. Some of its properties and numerous new features of the resulting algebra have since been analyzed and discussed \cite{mendes1996quantum,mendes2000geometry,das2006physics,ahluwalia2008ppa,ahluwaliakhalilova2005:2,kong2007quantum,kong2008deformed,kong2009ads_5,kong2009poincare,mendes2012deformation,mendes2015extended} , and a representation in terms of the Clifford algebra was provided by Gresnigt et.al. \cite{gresnigt2007sph}.

As an example, the algebra of Galilean relativity is unstable. Its stabilization requires the introduction of one deformation parameter $c$ (or rather $\frac{1}{c}$) where $c$ is physically identified with the speed of light and provides a maximum value for speeds. The resulting stable algebra is the Lorentz algebra and is isomorphic for all non-zero values of $c$. The process of stabilization therefore provides us with a generalized set of symmetries encoded within a stable algebra.\footnote{Similarly, the transition from the unstable Poisson algebra to the stable Heisenberg algebra is achieved through a deformation dependent on a new parameter which in this case is identified with $\hbar$.}

The deformation parameters that arise in the stabilization of an algebra correspond to new invariant scales. These invariant scales constrain the permissible values that certain kinematic variables can take and therefore restricts the theory. We note however that the numerical values of these invariant scales are not obtained via algebraic considerations but must be determined in the laboratory. Lie-type deformations therefore provide a systematic method of introducing additional invariant scales. Stability is both a physically sensible requirement in the sense that it provides an indication of the robustness of a theory, as well as mathematically unambiguous in the sense that the stable algebra is unique up to isomorphism. 

\section{Deformation of Galilean relativity to special relativity}
We here present a brief review of how last century's revolution from Galilean relativity to special relativity may be considered as a Lie-algebraic deformation. For the technical details the reader is referred to Chryssomalakos and Okon \cite{chryssomalakos2004:1}.

The metric of Galilean relativity is given by $d\mathbf{r}^2=d\mathbf{x}^2=dx^2+dy^2+dz^2$ with associated isometry group $ISO(3)=T_3\otimes_sSO(3)$. Here $T_3$ is the three-dimensional abelian translation group, $\otimes_s$ the semidirect product, and $SO(3)$ is the ordinary three-dimensional rotation group.

The symmetry transformations of Galilean relativity are rotations $R_i$ in three dimensions together with three Galilean boosts $N_i$ corresponding to (special) translations in three dimensions. The Galilean boosts commute and the Lie commutators encoding the symmetries of Galilean relativity are:
\begin{eqnarray}
\left[ J_{ij},J_{kl}\right] &=&-i(\delta_{jk}J_{il}+\delta_{il}J_{jk}-\delta_{ik}J_{jl}-\delta_{jl}J_{ik}),\\
\left[ J_{ij},N_k\right] &=&-i(\delta_{jk} N_i-\delta_{ik} N_j),\\
\left[ N_i,N_j\right] &=&0.
\end{eqnarray}

Galilean relativity imposes no constraints on any kinematic variables. In particular the speeds $v_i$, which act as the parameter in the Galilean boosts, are unbounded and take their values on the coset space $v^i \in ISO(3)/SO(3)=\mathrm{R}^3$.

With the discoveries of Einstein, Galilean relativity should now be restricted by incorporating an invariant maximum velocity $c$ into the theory. This means that no longer are $v^i\in \mathrm{R}^3$ because velocities are now bounded above by $c$. The introduction of the invariant scale deforms the commutator between Galilean boosts which previously commuted, 
\begin{eqnarray}
\left[ N_i,N_j\right] =0 \rightarrow \left[ N_i,N_j\right] =\frac{-i}{c^2}M_{ij}.
\end{eqnarray} 

Wishing to keep the theory linear, time $t$ is now promoted from an external parameter to a fourth dimension. This is made possible through the new invariant scale $c$ which allows time and space to be expressed in common units. The underlying physical space is changed from three-dimensional space to four-dimensional Minkowski spacetime with four vector
\begin{equation}
x_\mu=(x_0,x_1,x_2,x_3)=(ct,x,y,z)\equiv(ct,\mathbf{x}).
\end{equation}
The invariant $c$ likewise allows energy and momentum to be expressed in common units, leading to an associated four vector in momentum space
\begin{equation}
p_\mu=(p_0,p_1,p_2,p_3)=(c^{-1}E,p_x,p_y,p_z)\equiv(c^{-1}E,\mathbf{p}).
\end{equation}
Thus by introducing an invariant dimensionless scale $c$ into the position sector of the theory, its reciprocal is  simultaneously introduced in the momentum sector of the theory. 

We require that $ct\geq|\mathbf{x}|$ and it follows for some quantity $s_1$ that 
\begin{equation}
s_1^2+\mathbf{x}^2=c^2t^2,
\end{equation}
Rearranging and writing this in infinitesimal form we get the familiar metric of Minkowski spacetime
\begin{equation}
ds_1^2=c^2dt^2-d\mathbf{x}^2=c^2dt^2-dx^2-dy^2-dz^2.
\end{equation}

Identifying $N_i=J_{0i}$ above, we obtain a linear realization of the transformation group. The resulting symmetry group is the Lorentz group $SO(1,3)$, satisfying the commutation relation
\begin{eqnarray}
\left[ J_{\mu\nu},J_{\rho\sigma}\right] &=&-i(\eta_{\nu\rho}J_{\mu\sigma}+\eta_{\nu\sigma}J_{\nu\rho}-\eta_{\mu\rho}J_{\nu\sigma}-\eta_{\nu\sigma}J_{\mu\rho}).
\end{eqnarray}
This group is semisimple ad thus stable. The velocity $v^i$ is now constrained to take values on the coset $v^i\in SO(1,3)/SO(3)$. In the limit that $c\rightarrow \infty$ we recover 
\begin{eqnarray}
\left[ J_{0i},J_{0j}\right] &=&0,
\end{eqnarray}
and the algebra again splits to become that of Galilean relativity
\begin{eqnarray}
SO(1,3)\vert_{c\rightarrow \infty} \rightarrow SO(3)\otimes _s R^3.
\end{eqnarray}
Mathematically, taking the limit $c\rightarrow \infty$ corresponds to a Inonu-Wigner contraction. Such a contraction is essentially the reverse process of a Lie-type deformation. A contraction limit represents an approximate description (Galilean relativity) of the full stable symmetry (special relativity).

The Lorentz group above is semisimple and hence stable. To obtain all the isometries however we must also add four commuting translations $P_{\mu}$ onto our four-dimensional manifold. The result is the Poincar\'e algebra 
\begin{eqnarray}
\left[ J_{\mu\nu},J_{\rho\sigma}\right] &=&-i(\eta_{\nu\rho}J_{\mu\sigma}+\eta_{\nu\sigma}J_{\nu\rho}-\eta_{\mu\rho}J_{\nu\sigma}-\eta_{\nu\sigma}J_{\mu\rho}),\\
\left[ J_{\mu\nu},P_{\rho}\right] &=& -i(\eta_{\mu\rho}P_{\nu}-\eta_{\nu\rho}P_{\mu}),\\
\left[ P_{\mu},P_{\nu}\right] &=& 0.
\end{eqnarray}
with associated Poincar\'e isometry group $ISO(1,3)=T_4\otimes _s SO(1,3)$\footnote{Strictly speaking the full Poincar\'e group is $T_4\otimes_sO(1,3)$ but we are ignoring the discontinuous space and time inversion symmetries in this paper.}. The inclusion of a four-dimensional translation group $T_4$ destabilizes the isometry group which is now again unstable. Since the isometry group has an unstable algebra, we conclude that further kinematic contraints in addition to $c$ must be imposed. This means further deforming the isometry group (in this case the Poincar\'e group of special relativity).

At this stage we pause and reflect briefly on the physical consequences deforming Galilean relativity has had on our theory. First, the deformation parameter in a Lie-type deformation corresponds to a physical invariant scale, in this case $c$. Second, a Lie-type deformation changes the underlying Lie algebra and thus the symmetries. Finally, wishing to keep the theory linear, the deformation has led to a change in the underlying physical space (from three dimensional space to four dimensional spacetime) and associated metric. The deformed metric depends explicitly on the new invariant scale. One can expect similar changes to the theory and underlying spacetime as further deformations are considered.

\section{Deforming special relativity}

From the previous section one notes that stabilizing the isometry group of Galilean relativity gives the stable Lorentz algebra. However, the full set of isometries of Minkowski spacetime also includes the four spacetime translations. Including these renders the full isometry group unstable again. We find ourselves in the same situation as with Galilean relativity (an isometry group consisting of a semidirect product of a stable component and a translation group) but with a larger isometry group. Repeating this process of stabilizing and extending by translations will introduce additional invariant scales but will in general not resolve the underlying issue. However, when an invariant length scale is introduced, the resulting spacetime is restricted to a hypersurface of one fewer dimensions. Translations are then no longer admissible isometries and this deformation process terminates.  

In the present paper we focus therefore on deforming the Poincar\'e algebra once more only via the introduction of a quantum length scale $\ell$. This terminates the sequence of deformations. Notice that in theory it is possible to introduce multiple additional invariant scales \cite{das2006physics,chryssomalakos2004:1,kowalskiglikman2004tsr,kong2008deformed}. The key point here is that the deformation process continues until one is restricted to a hypersurface and translations are no longer permissible isometries. We here investigate the simplest case consisting of one additional deformation. By imposing $c$ and $\ell$ as invariant scales we restrict the range of all the kinematic variables. On the one hand we are enlarging the theory via the introduction of additional constants and deforming Lie algebras whereas on the other hand we are restricting the Galilean theory by constraining the range of kinematic variables. 

\subsection{An invariant minimal length scale}

The introduction of an invariant length scale ends the iterative process of deforming an isometry group and adding on a translation group. However, so far we have provided no reason as to why a length scale, and in particular a \textit{minimal} length scale is a sensible constraint. 

Quantum gravity is likely to introduce phenomenology that differs from that of general relativity. There should therefore exist some scale that marks the threshold for such new phenomenology. The principle of relativity then requires that this scale should be invariant to guarantee the same phenomenology for all inertial observers. Combining gravity ($G$), relativity ($c$), and quantum mechanics ($\hbar$) gives rise to the Planck length $\ell_p=\sqrt{\hbar G/c^3}$, Planck time $t_p=\sqrt{\hbar G/c^5}$ and Planck mass $m_p=\sqrt{\hbar/Gc}$. These fundamental scales can be thought of as marking the scale at which the description of spacetime provided by special and general relativity breaks down and new phenomenology can be expected. One of these scales should therefore be an invariant. This is the motivation behind Doubly Special Relativity (DSR) which in addition to $c$ introduces the Planck length $\ell_P$ or mass $m_P$ as an additional invariant \cite{amelino2002relativity,amelino2002doubly2,kowalski2001observer,kowalski2005introduction}. The disadvantage of DSR theories however is that it lacks the systematic approach offered within a Lie-type deformation framework.

Alternatively, consider the thought-experiment to probe spacetime at spatial resolutions around the Planck length $\ell_P$. Doing so creates a Planck mass blackhole carrying a temperature of $T\approx 10^30$ K and with an evaporation time of $t\approx10^{-40}$ sec. This formation and evaporation of the blackhole places a lower bound on the spatial resolution with which spacetime may be probed. The above thought-experiment suggests that $\ell_p$ will inevitably play an important role in defining quantum spacetime \cite{ahluwalia2008ppa}.

Many attempts to explain some of the peculiar features of the SM rely on replacing the point particles with extended objects that are allowed to carry topological information. Assuming the size of these objects to be on the order of $\ell$, this length scale then marks the boundary between the internal structure of a particle and the surrounding spacetime. An external observer cannot maintain a distinction of being an external observer at distance smaller than $\ell$. This means that spacetime isometries are only valid at scales greater than $\ell$. 

\subsection{Relativity with minimal length scale}

Introducing additional invariant scales leads to changes in the spacetime metric, underlying physical space, and the Lie algebra encoding the isometries. Associated with a minimal length scale $\ell$ in spacetime is associated a maximum momentum in momentum space. Furthermore, together with $c$ we also obtain implicitly an invariant mass, time, and energy \footnote{Strictly speaking this is only true if one assumed $\hbar$, an issue that has been discussed by Kong \cite{kong2008deformed}. $\hbar$ can itself be considered as the deformation parameter in the deformation of the Poisson algebra to the Heisenberg algebra. We do not focus in the present paper on the emergence of $\hbar$ from a deformation and thus simply assume it.}. 

The deformation from Galilean relativity to special relativity was dependent on restricting the range of the parameter of the special translations (that is; we restricted the speed in the Galilean boosts). Similarly we may now restrict the Poincar\'e translations by introducing a maximum momentum $\ell^{-1}$. We modify our theory the same way we modified the original $T_3\otimes_sSO(3)$ theory. This means we consider a five-dimensional manifold and write down the five-vector
\begin{equation}
x_A=(x_0,x_1,x_2,x_3,x_4)=(ct,x,y,z,\ell\rho)=(x_\mu,\ell\rho).
\end{equation}
Here $\rho$ in natural units is a dimensionless parameter. In writing down a five-vector in the position sector of the theory we also automatically have a five-vector in the momentum sector of the theory:
\begin{equation}
p_A=(p_0,p_1,p_2,p_3)=(c^{-1}E,p_x,p_y,p_z,\ell^{-1}\lambda)=(p_\mu,\ell^{-1}\lambda).
\end{equation}
We might now wonder about the metric on the position sector. Requiring $|x_\mu x^\mu|\geq\ell_p^2\rho^2$, or, $x_\mu x^\mu\geq -\ell^2\rho^2$ implies that there exists an $s$ such that $s^2-\ell^2\rho^2=x_\mu x^\mu$, or, in infinitesimal form,
\begin{equation}
ds^2=dx_\mu dx^\mu +\ell_p^2d\rho^2=c^2dt^2-dx^2-dy^2-dz^2+\ell^2d\rho^2.
\end{equation}

In terms of the Lie algebra, the introduction of an invariant length $\ell$ means the commutators are modified, with spacetime translations no longer commuting
\begin{eqnarray}
\left[ P_{\mu},P_{\nu}\right] = 0 \rightarrow \left[ P_{\mu},P_{\nu}\right] = \frac{-i}{\ell ^2}J_{\mu\nu}.
\end{eqnarray}
The other Poincar\'e commutators remain unaffected. Identifying $P_{\mu}=J_{\mu 4}$ the three commutation relations in four dimensions simplify to a single commutation relation in five dimensions
\begin{eqnarray}
\left[ J_{AB},J_{CD}\right] &=&-i(\eta_{BC}J_{AD}+\eta_{AD}J_{BC}-\eta_{AC}J_{BD}-\eta_{BD}J_{AC}).
\end{eqnarray}
The symmetry group is now $SO(2,3)$ with the momenta restricted to the coset space $P_\mu=J_{\mu 4} \in SO(2,3)/SO(1,3)$. This corresponds to the anti de Sitter group.

At this point we have a constraint that restricts the physically relevant space to a four-dimensional hypersurface of the original five-dimensional space. This means that translations are no longer isometries. We may no longer add a translation group to extend the isometry group to $T_5\otimes_sSO(2,3)$. The induced metric on this four-dimensional hypersurface has isometry group $SO(2,3)$. This group is semisimple and thus stable. No further deformations are required \footnote{As noted earlier, there is in theory nothing preventing one from deforming the Poincar\'e algebra with respect to other invariant scales. Such deformations will in general not restrict the physical space to a hypersurface.}

\subsection{Quantum spacetime as anti de Sitter space}

Extending special relativity beyond Einstein (Poincar\'e) through the introduction of an invariant minimal length scale, the isometry group becomes $SO(2,3)$ which is the Anti de Sitter group corresponding to Anti de Sitter space which carries a negative curvature. Cosmological experiments suggest that the overall curvature of the spacetime manifold is positive. This corresponds to a positive cosmological constant and to a de Sitter universe. We note however that there is no contradiction with our work here. The scale we introduced is a minimal length scale. We are therefore restricting ourselves to the quantum scale and argue that at the $\ell$ scale, spacetime looks like Anti de Sitter. The large scale structure of the universe may still carry positive curvature. This presents a rather interesting view of spacetime in which on the quantum scale, spacetime is negatively curved, whereas on large scale the universe carries positive curvature. We do not however discuss this point in detail in this paper.

Deforming the isometries of Minkowski spacetime to those of Anti de Sitter space has important implications to the description of particles. Elementary particles in Minkowski spacetime are associated with unitary irreducible representations of the Poincar\'e group (the isometry group of Minkowski spacetime). We have here deformed the underlying spacetime from Minkowski to Anti de Sitter along with the associated isometry groups $ISO(1,3)$ to $SO(2,3)$. In our deformed relativity, elementary particles should therefore be associated with the unitary irreducible representations not of the Poincar\'e group, but of the Anti de Sitter group. 

The representations of the Anti de Sitter group, like the Poincar\'e group, admit a positive minimum energy. This means that representations of this group naturally lends themselves to a particle interpretation. This is not the case for de Sitter space with positive curvature. This has caused confusion given the observations of a positive cosmological constant. In the present scheme however we naturally arrive at Anti de Sitter space for quantum spacetime which is still consistent with a large scale positive cosmological constant. 

The representations of the Anti de Sitter group $SO(2,3)$ have been studied extensively in the literature. The most fundamental irreducible representations of this group were first discovered by Dirac \cite{dirac1963remarkable}, and are called the singleton representations. The physics associated with these singleton representations has been studied most notably by Flato and Fronsdal, see \cite{flato1999singleton} and references therein. In Anti de Sitter space, massless particles are composed of two singletons. These singletons are naturally confined (in a kinematic sense) which has also led to the question of whether perhaps singletons take the role of quarks \cite{flato1986quarks}. 

Compatibility between the singleton representations and quantum electrodynamics (QED) was demonstrated by Flato and Fronsdal \cite{flato1981quantum}. What is fascinating here, is that the resulting theory is a topological field theory, in agreement with our assumption that elementary particles should not be thought of as pointlike, but rather as being topological in nature. It then seems natural to speculate, as Sternheimer, Flato, and Fronsdal have done on several occasions \cite{flato1986quarks,flato1999singletons} that elementary particles are composed of multiple (anti-) singletons. We see a similarity here with various preon models \cite{harari1979schematic,shupe1979composite,bilson2005topological}. Although not a focus of the present paper, it would be interesting to investigate this possible connection between singletons and preons in more depth. 

The group representation theory of $SO(2,3)$ suggests that the unitary irreducible representations are composed of two or more degeneratre UIRs of the covering of $SO(2,3)$. These degenerate UIRs are the singletons that were first discovered by Dirac (see \cite{sternheimer2007geometry} page 294). 

The two singleton representations, named \textit{Di} and \textit{Rac} are given by
\begin{eqnarray}
Rac=D(\frac{1}{2}, 0),\qquad Di=D(1,\frac{1}{2}).
\end{eqnarray}
These singleton representations have the interesting property that a direct product of two positive energy singletons reduces to a sum of massless representations of $SO(2,3)$ as follows \cite{flato1978one}
\begin{eqnarray}
Rac\otimes Rac&=&\oplus_{s=0,1,...}D(s+1,s),\\
Rac\otimes Di&=&\oplus_{2s=1,3,...}D(s+1,s),\\
Di\otimes Di&=&\oplus_{s=1,2,...}D(s+1,s)\oplus D(2,0).
\end{eqnarray}

The $Di$ and $Rac$ themselves do not have contractions to representations of the Poincar\'e group and so at the flat space limits, the singletons reduce to vacua. The physics of these singletons goes beyond the Minkowski space limit and has no analogue. Particularly, in the case where space carries nonzero curvature, massless particles can be considered as composite objects. This ceases to be true in flat space. This means that massless particles in (anti) de Sitter spacetime differ from massless particles in Minkowski spacetime.

\section{Composite and extended topological SM particles}

Several approaches to a more foundational model of particle physics have been considered. These include preon models in which the currently elementary particles are considered to be composite. Perhaps the most famous of these is the Harari-Shupe model \cite{harari1979schematic,shupe1979composite}. More recently Bilson-Thompson \cite{bilson2005topological} showed that the simplest braids consisting of three ribbons and two crossings map precisely to the first generation of SM leptons and quarks. It was later demonstrated that these braids may be embedded within spin networks, which makes it compatible with background independent theories of quantum gravity, for example loop quantum gravity \cite{bilson2007quantum,bilson2009particle,bilson2008particle,he2008c,he2008conserved,hackett2007locality}. 

Apart from preon models, Jehle in 1971 suggested a model of particles based on quantised flux tubes \cite{jehle1971relationship,jehle1972flux,jehle1975flux}. The stability of such solitons was demonstrated for the case of the unknot and the trefoil knot by Faddeev and Niemi \cite{faddeev1996knots,faddeev1997stable} and later for a wider class of knots by Kobayashi and Nitta \cite{kobayashi2014torus}. A similar model of particles in terms of trefoil knots was considered by Finkelstein \cite{finkelstein2005knot,finkelstein2007elementary,finkelstein2010solitonic,finkelstein2012slq,finkelstein2006masses}.

A common theme among these approaches is to replace the notion of a point particle with the notion of an extended\footnote{and likely deformable, depending upon the substance of which they are made. For an interesting discussion on the nature of the 'stuff' that makes up matter, the reader is directed to a recent paper by van der Mark \cite{Mark2015a}.} particle that carries topological information. The numerous interesting and encouraging results obtained in the above cited works suggest this approach is worth developing further. At the same time, stability considerations lead us to question the validity of the unstable Poincar\'e algebra that underlies the spacetime symmetries of the SM.

We therefore supplement the hypothesis that elementary particles be extended objects carrying topological information with the requirement of Lie-algebraic stability. That is, these extended objects should live in anti de Sitter spacetime rather than Minkowski spacetime. The topological information carried by elementary particles should than be compatible with the topological properties of the spacetime in which they live. 

\section{The possible topological connection between internal and space-time symmetries}
The SM internal symmetries present one of the most controversial aspects of the SM as the fundamental origin of the internal symmetries is not known. A natural question to ask is if there exists any connection between the internal symmetries and the external space-time symmetries. This question was asked many times in the 1960's in particular during which time several no-go theorems emerged. These theorems seemingly proved that the connection between internal and spacetime symmetries (specifically Poincar\'e symmetry) could only be trivial, in the sense of a direct product of symmetry groups. 


\subsection{No-go theorems}
 
The first no-go theorem is accredited to L. O'Raifeartaigh who in 1965 claimed that only a trivial combination of the internal and spacetime symmetries is possible \cite{o1965mass}. The proof argues that the nilpotency of the momentum generators $P_{\mu}$ in the Poincar\'e algebra or any finite-dimensional Lie algebra containing the Poincar\'e algebra forbids a discrete mass spectrum. Therefore, requiring a discrete mass spectrum means that the connection between the internal and spacetime symmetries must be a direct product. The result was however challenged by Flato and Sternheimer who showed that the proof contained an error \cite{flato1965remarks}. They further provided several counter examples \cite{flato1966local,flato1969poincare}. 

A more sophisticated no-go theorem was formulated two years later by Coleman and Mandula \cite{coleman1967all} based on the symmetries of the $S$-Matrix. But again it was shown \cite{flato1969poincare} that the proof was incomplete and contained a hypothesis that presupposed the result that was proved.

Given the above counterexamples, Sternheimer has argued that one cannot and should not rule out a priori any nontrivial relation between the internal and spacetime symmetries \cite{sternheimer2007geometry}. We agree. The no-go theorem of L. O'Raifeartaigh only applies to any finite-dimensional Lie algebra that contains the Poincar\'e algebra. Our deformation of special relativity to anti de Sitter relativity means this condition is no longer satisfied. In particular the nilpotency of the generators of spacetime translations is lost with the introduction of a minimal length scale.

\subsection{Extended particles and no-go theorems}

Assuming that elementary particles carry topological information one might hope that the internal symmetries of these particles can be described as emerging from their topological properties. In such a case, it is physically sensible to require the topology exhibited by particles to be compatible with the underlying topological properties of spacetime in which these particles live. 


If the particle is point-like with no spatial internal structure then its internal symmetries are not spacetime related and so one naturally expresses the theory in symbolic form as
\begin{equation}
\mathrm{External}\times\mathrm{Internal}:H\longrightarrow H,
\end{equation}
where $H$ is a Hilbert space, preserving the distinction between external spacetime symmetries and internal non-spacetime symmetries. That is, no-go theorems essentially say that one needs to preserve a distinction between a particle and its surroundings. 

Replacing point particle by extended topological objects of characteristic size $\ell$, means that the scale $\ell$ defines a demarcation between external spacetime observers and the internal space of a particle. This demarcation is important as it distinguishes between those regions of space accessible and inaccessible to external observers. We therefore likewise conclude that 
\begin{equation}
\mathrm{External}\times\mathrm{Internal}:H\longrightarrow H,
\end{equation}
however now, there is no a priori reason to suspect that the internal symmetries are unrelated to the external spacetime symmetries. If particles are extended objects, that means that they deform the spacetime around them, and vice versa. It then makes sense that the internal symmetries are related to the spacetime symmetries whilst concurrently it tells us we need to pull them out and add them in separately in a trivial way as suggested by the no-go theorems.

In the case where particles are considered to be extended, no-go theorems then seem to lend themselves to a natural interpretation which says that one must keep a distinction between the spacetime internal to a particle and the spacetime external to a particle, with external observers being restricted to the latter. 

Additionally, one should maintain a basic consistency between the internal structure of particles and the topological properties of spacetime, in that the two must be compatible. 

\subsection{Electroweak symmetry from quantum-scale anti de Sitter spacetime}

Compatibility between particles and their surrounding spacetime in which they live means that the permitted structures are controlled by the topology of the underlying spacetime (and its isometries). The topological information of spacetime is contained within its homotopy groups, most importantly its fundamental group. Thus, the homotopy groups of spacetime then dictate the allowed internal structures.

The homotopy groups of a group $G$ are the same as those of its maximal compact subgroup $K$. In our case this means that the internal topology of a particle is given by the maximal compact subgroup of $SO(2,3)$. This is the direct product group $SO(2)\times SO(3)$. Consistency with quantum mechanics means we want unitary groups instead giving $U(1)\times SU(2)$, the electroweak gauge group, which is locally isomorphic to the anti de Sitter maximal compact subgroup.

The need to distinguish between the spacetime internal to a particle and the spacetime surrounding it means that the internal symmetry group (corresponding to the maximal compact subgroup of spacetime isometries) must be written distinctly from the external observer's isometry group (the isometries of the spacetime metric) in the form dictated by no-go theorems. This we believe provides a physical reason as to why no-go theorems exist. The resulting theory then takes the form
\begin{eqnarray}
SO(2,3)\times SU(2)\times U(1):H \longrightarrow H.
\end{eqnarray}

\subsection{Strong interactions}

Our approach has provided an explanation for the origins of electroweak symmetry in SM particles without any reference to the strong force. This is what we should expect. The fact that all fermions experience the electroweak force is reflected in the general nature of our approach, which should therefore be universally applicable. The strong force on the other hand is not felt by all SM fermions. Furthermore, strongly interacting particles are bound. The nature of the strong force is therefore very different and we should not expect it to arise from the same considerations. A complete and satisfactory picture should however explain the possible origin of the strong force. We here suggest two possibilities but leave it for a future opportunity to investigate these in depth.

One possibility is that the singleton representations adequately describe the strong interaction. This possibility has also been considered by Flato and Fronsdal \cite{flato1986quarks}. 

Another possibility is that the strong force arises from a further deformation. The present isometry group is semi-simple, and thus no further deformations of Lie-type are possible. Additional deformations should then be of a different type. One possible candidate is quantum Hopf deformations (or simply $q$-deformations) which give quantum Hopf algebras otherwise known as quantum groups. To obtain the full structure of spacetime along with the internal symmetries of the SM (and beyond), one might deform from $ISO(1,3)$ to $SO(2,3)$ and subsequently to $SO_q(2,3)$. Such an approach seems promising given the relation between quantum groups and knot invariants. This also is consistent with particles being fundamentally topological in nature. The maximal compact subgroup is likewise $q$-deformed.

Finkelstein has shown that there exists a one-to-one correspondence between the state labels of trefoil representations of $SU_q(2)$ and the quantum numbers of the electroweak theory, and that the strong force can be described in this context by extending the theory to $SL_q(2)$   \cite{finkelstein2005knot,finkelstein2007elementary,finkelstein2003note}. Furthermore, $q$-deformation contains additional information which he argues characterizes the soliton nature of the SM fermions. 

The preservation of SM physics under $q$-deformations, together with the additional information provided, are encouraging indications that the present theory will need to be $q$-deformed. 

\section{Discussion and conclusion}

In this paper we have argued that Lie-type stability plays an important role in finding the appropriate isometries of a physical theory. An unstable isometry group should be deformed, via the introduction of additional invariant scales, into a stable one. The resulting stable symmetry is more widely applicable. The introduction of invariant scales restricts the ranges of kinematic variables. The spacetime symmetries of the SM and high energy physics are encoded in the unstable Poincar\'e algebra. A natural path toward a more robust theory of high energy physics is to replace this underlying unstable algebra with a deformed stable one. 

The minimal deformation of Poincar\'e isometries to a stable isometry group is through the inclusion of an invariant length scale $\ell$. Together, $c$ and $\ell$ are sufficient to constrain the ranges of all kinematic variables. The introduction of a minimal length scale $\ell$ prohibits the inclusion of translations as admissible isometries and thereby prevents the destabilization of the isometry group (as is the case going from Lorentz symmetry to Poincar\'e symmetry). The introduction of a minimal length scale $\ell$ on the order of particle sizes, deforms (on the quantum scale) Minkowski spacetime into anti de Sitter spacetime (AdS) with associated isometry group $SO(2,3)$. 


The concept of introducing additional invariant scales into relativity is not new. Amelino-Camelia argued for the introduction of an invariant length scale on the order of the Planck length in order to mark the scale at which quantum gravitational phenomenology may be expected \cite{amelino2002doubly2,amelino2002relativity,amelino2001testable}. The resulting class of theories are referred to as Double Special Relativity (DSR) \cite{amelino2003phenomenology,kowalski2005introduction,kowalski2001observer,kowalski2003doubly,kowalski2002doubly,magueijo2002lorentz,schutzhold308049pds,lukierski2002doubly,girelli2005deformed}. Kowalski-Glikman and Smolin later introduced a third invariant scale (the cosmological constant) thereby extending DSR to TSR (Triply Special Relativity) \cite{kowalskiglikman2004tsr}. These deformed theories of relativity generally introduce non-linearity into the theory (i.e. the Lie-algebraic framework is abandoned). The advantage of Lie-type deformations is that they can be handled systematically without introducing non-linearity. Indeed, Chryssomalakos and Okon have shown that through a suitable redefinition of the generators, TSR can be brought into linear form \cite{chryssomalakos2004:2}. There seems little reason therefore to deviate from the linear framework provided by Lie algebras. 

Our requirement of stability not only provides a logical step towards a theory beyond the SM, it also explains why the SM is so successful despite the instability of its underlying algebraic structure. It is readily seen that the Lorentz sector remains undeformed while modifications to the theory are introduced via the invariant scale $\ell$. The magnitude of the deformation scale (which we assume to be very small, on the order of the size of elementary particles) means the deformations are inaccessible at low energies\footnote{Parenthetically we point out that $SO(2,4)/Z_2$ is isomorphic to $SU(2,2)/Z_4$. The maximal compact subgroup of $SU(2,2)$ is $S(U(2)\times U(2))$ whose representations depend on $(j_1,j_2)$, or equivalently, the irreducible representations of $SU(2)\times SU(2)$. This offers an explanation as to why the Lorentz sector survives in the bigger group.}. 

Several topological models of particles have been developed in attempts to explain some of the unexplained features of the SM. These include preon models, most famously the Harari-Shupe model, and more recently the Helon model, as well as knotted flux tube models of elementary particles. We have argued that replacing the point particles of the SM with extended topological objects leads to the requirement that the topology of the particles must be consistent with that of the underlying spacetime, which from stability considerations is anti de Sitter. The homotopy groups of a group, containing its essential topological information, are the same of those of its maximal compact subgroup. Together with the requirements of unitarity from Wigner's general quantum mechanics symmetry representation theorem, this directly gives us the electroweak sector of the SM. 

Our approach is compatible with no-go theorems and provides a physical motivation for them. One needs to keep a distinction between the spacetime internal to a particle and the spacetime surrounding the particle, with external observers being restricted to the latter. At the same time, the internal topological structure must be consistent with the topology of the external spacetime. The origin of the electroweak symmetries is therefore the maximal compact subgroup of spacetime, and the required distinction between the spacetime internal and external to particles means that these internal symmetries must be added separately to the spacetime symmetries in the trivial direct product sense implied by no-go theorems.

\.Zenczykowski discovered a deep connection between the symmetries of non-relativistic phase space and the internal symmetry group of the SM. His approach depends on mixing space and momentum which requires the introduction of a mass scale. Such a mass scale is obtained by supplementing $c$ and $\hbar$ with a new invariant length scale \cite{zenczykowski2008harari,zenczykowski2007space,zenczykowski2007structure}. Our work shows that \.Zenczykowski's required length scale arises naturally in the minimal deformation of the Poincar\'e algebra. It would be interesting to further investigate the possible connection between the present work and the work of \.Zenczykowski.

Our theory is not yet complete in its present form with a consideration of the strong force still lacking. One proposal worth investigating is to $q$-deform the isometry group $SO(2,3)$ to $SO_q(2,3)$. The electroweak sector under $q$-deformation can be described by $SU_q(2)$ as explained by Finkelstein. This quantum Hopf algebra is the algebra of oriented knots and suggests that particles can be considered as non-pointlike quantum solitons. This is a strong indication that the present work is consistent with the models of Jehle \cite{jehle1971relationship,jehle1972flux}, Faddeev \cite{faddeev1996knots,faddeev1997stable}, and Finkelstein \cite{finkelstein2005knot,finkelstein2007elementary,finkelstein2010solitonic,finkelstein2012slq,finkelstein2006masses}. We wish to address these issues more deeply in a future paper.


\subsubsection*{Acknowledgments}
The authors wish to thank Otto Kong for stimulating discussions and Ben Martin for his insightful comments on a draft version of this paper. This work is partially supported by the research grant RDF-14-03-13 from XJTLU.

\bibliography{SPHA}  
\bibliographystyle{unsrt}  

\end{document}